\documentstyle[prd,aps]{revtex}
\begin{document}
\draft
\title{Renormalization of nonequilibrium dynamics in FRW cosmology}
\author{J\"urgen Baacke
\footnote{Electronic address:~baacke@physik.uni-dortmund.de},
Katrin Heitmann \footnote{
Electronic address:~heitmann@hal1.physik.uni-dortmund.de},
and Carsten P\"atzold \footnote{
Electronic address:~paetzold@hal1.physik.uni-dortmund.de}
}
\address{Institut f\"ur Physik, Universit\"at Dortmund\\
         D - 44221 Dortmund, Germany}
\date{June 10, 1997}
\maketitle
\begin{abstract}
We derive the renormalized nonequilibrium
 equations of motion for a scalar field and its quantum 
back reaction 
in a conformally flat Friedmann-Robertson-Walker
universe. We use a fully covariant formalism proposed by us recently 
for handling numerically and analytically nonequilibrium
dynamics in one-loop approximation. 
The system is assumed to be in a conformal vacuum state
initially. We use dimensional
regularization; we find that the counter terms can be chosen independent
of the initial conditions though the divergent leading order
graphs do depend on them.
\end{abstract}


\section{Introduction}
Nonequilibrium processes in cosmology have been considered
recently by various authors. The main interest has been
centered around the possible inflationary period of the universe
(see e.g. \cite{Abbott:1986,Kolb:1990,Linde:1990})
and the subsequent reheating \cite{Dolgov:1982,Abbott:1982}.
It has already been found by considering the parametric
resonance \cite{Kofman:1994}
associated with oscillations of the inflaton field
and by exact computations including back reaction
both in Minkowski space 
\cite{Boyanovsky:1994} and in an expanding universe
\cite{Boyanovsky:1997a,Kofman:1997} that
the time-dependent inflaton field produces particles or
classical fluctuations preferentially
at low momenta and not in a distribution corresponding to
thermal equilibrium. 
 This process of particle
production has therefore been termed preheating
\cite{Kofman:1994,Shtanov:1995}.

The equations of motion for nonequilibrium systems
have been presented by various authors 
\cite{Jordan:1986,Calzetta:1987,Calzetta:1988} using
the CTP formalism introduced by Schwinger \cite{Schwinger:1961} 
and Keldysh \cite{Keldysh:1964}. Their application to inflation
within a conformally flat FRW universe has been
initiated by Ringwald 
\cite{Ringwald:1988,Ringwald:1987a,Ringwald:1987b};
they have been recently implemented numerically
in \cite{Boyanovsky:1997b} and \cite{Ramsey:1997}.
Similar computations have been 
performed recently in configuration space
for the case that the fluctuations are
treated as classical ones 
\cite{Khlebnikov:1996,Khlebnikov:1997}. Again fluctuations with rather
low momenta are strongly excited, thus justifying the
classical approximation.
Apart form such exact numerical computations there exist
also various analyses based on analytical approximations to the
solution of the mode equations 
\cite{Kofman:1997,Son:1996,Greene:1997a,Kofman:1996,Kaiser:1996,Kaiser:1997}.

We have proposed recently a computation scheme
for nonequilibrium dynamics which has
two aspects: on the one hand it separates cleanly the divergent and the
finite contributions of the quantum fluctuations in one-loop
approximation, and so is attractive for numerical computations.
On the other hand it leads, by the analysis of the divergent
leading order contributions, to a simple formulation of the
renormalized equations of motion. The fact that the divergent
contributions are removed from the numerical computation
allows a free choice of regularization. We have
chosen dimensional regularization since it 
is easier to handle in the presence of quartic and
quadratic divergencies than 
e.g. Pauli-Villars regularization.This scheme has been
applied and  implemented numerically for scalar fields
\cite{Baacke:1997a}, fermion fields 
and for the SU(2) Higgs model \cite{Baacke:1997b}.
We will present here the renormalized equations of motion
for a scalar field in  flat Friedmann-Robertson-Walker (FRW)
cosmology.    

We will not attempt here a numerical implementation.
There exist already several numerical analyses based on
various assumptions on the initial conditions and on the 
parameters of the theory.
We do not expect that our formulation will lead to major
differences in the  qualitative features of the results.
Nevertheless, we think that in the presence of
quartic and quadratic divergencies  it is important to formulate
the renormalization scheme in a way that is fully covariant
and independent of the initial conditions. It turns out
that the resulting formalism  represents at the same time 
a rather attractive numerical computation
scheme.

\section{FRW cosmology}
We consider the Friedmann-Robertson-Walker metric
with curvature parameter $k=0$, i. e. a spatially isotropic and flat
space-time.
The line-element is given in this case by
\begin{equation} 
\label{line}
{d}s^2={d}t^2-a^2(t){d}{\vec x} ^2\; .
\end{equation}
The time evolution of the $a(t)$ is governed
by Einstein's field equation
\begin{equation} \label{Einfield}
G_{\mu\nu}+\alpha  H^{(1)}_{\mu\nu}+\beta  H^{(2)}_{\mu\nu}
+\Lambda g_{\mu\nu}=
-\kappa
\langle T_{\mu\nu}\rangle\;,
\end{equation}
with $\kappa= 8\pi G$.
The Einstein curvature tensor $G_{\mu\nu}$ is given by
\begin{equation} 
G_{\mu\nu}=R_{\mu\nu}-\textstyle{\frac 1 2} g_{\mu\nu}R\; .
\end{equation}
The Ricci tensor and the Ricci scalar are defined as
\begin{eqnarray}
R_{\mu\nu}&=&R^{\lambda}_{\mu\nu\lambda}\; ,\\
R&=&g^{\mu\nu}R_{\mu\nu}\; ,
\end{eqnarray}
where
\begin{equation} 
R^{\lambda}_{\alpha\beta\gamma}=
\partial_{\gamma}\Gamma^{\lambda}_{\alpha\beta}
-\partial_{\alpha}\Gamma^{\lambda}_{\gamma\sigma}
\Gamma^{\sigma}_{\alpha\beta}
-\Gamma^{\lambda}_{\alpha\sigma}\Gamma^{\sigma}_{\gamma\beta}\; .
\end{equation}
The terms  $H_{\mu\nu}^{(1)}$ and $H_{\mu\nu}^{(2)}$ arise if
terms proportional to $R^2$ and $R^{\mu\nu}R_{\mu\nu}$ are
included into the Hilbert-Einstein action.
If space-time is conformally flat, these terms are related by
\begin{equation} 
 H^{(2)}_{\mu\nu}={\textstyle \frac{1}{3}} \mbox{ }H^{(1)}_{\mu\nu}\; ,
\end{equation}
so that we can set $\beta=0$ 
in (\ref{Einfield})
without loss of generality \cite{Birrell:1982}.
We also replace $H_{\mu\nu}^{(1)}$
by $H_{\mu\nu}$ in the following.

These terms are usually not considered in standard cosmology.
They are included here, as well as
the cosmological constant, only for the purpose of
renormalization; they will absorb divergencies of the energy momentum
tensor. So in principle they should appear on the right hand side
as counter terms for the energy momentum tensor;
they are related to the coefficients of these
counter terms
\footnote{See (\ref{Ewct}) for their precise definitions.} 
  by $\Lambda=\kappa\delta\tilde{\Lambda}$ and
$\alpha=\kappa\delta\tilde{\alpha}$. We will need 
a further counter term proportional to $G_{\mu\nu}$, which can
be considered as a wave
function renormalization for the gravitational field;
either we have to replace $G_{\mu\nu}$ by $Z G_{\mu\nu}$ on the
left hand side of the Einstein equations or to introduce a counter term 
$\delta\tilde{Z} G_{\mu\nu}$ for the energy momentum tensor.
Again both alternatives are related by $Z=1+\kappa \delta \tilde{Z}$.
Writing the renormalized equation (\ref{Einfield}) in the form
\begin{equation}
\left(\frac 1 \kappa +\delta \tilde{Z}\right)G_{\mu\nu}+\left(\tilde{\alpha}
+\delta\tilde{\alpha}\right)H_{\mu\nu}+\left(\tilde{\Lambda}+\delta\tilde{\Lambda}\right)g_{\mu\nu}=T_{\mu\nu}^{\rm{ ren}}
\end{equation}
we can identify $\delta\tilde{Z}$ as a correction to the gravitational coupling
via
$\kappa^{-1}\rightarrow \kappa^{-1}+\delta\tilde{Z}$.
As usual, we can reduce the Einstein field equations to an equation
for the time-time component and one for
the trace of $G_{\mu\nu}$, the Friedmann
equations
\begin{eqnarray}
\label{Friedtt}
G_{tt}+ \alpha H_{tt}+\Lambda &=&-\kappa T_{tt}\; ,\\
\label{Friedtr}
G_\mu^\mu+ \alpha H^{\mu}_\mu+4\Lambda &=&-\kappa T^{\mu}_{\mu}\; .
\end{eqnarray}
For the line element (\ref{line}) the various terms take the form
\cite{Birrell:1982}
\begin{eqnarray}
G_{tt}(t)&=&-3H^2(t)\; , \\
G_{\mu}^\mu(t)&=&-R(t) \; , \\
H_{tt}(t)&=&-6\left[H(t)\dot{R}(t)+H^2(t)R(t)
-\frac{1}{12} R^2(t)\right]\; , \\
H^{\mu}_\mu(t)&=&-6\left[\ddot{R}(t)+3H(t)\dot{R}(t)\right]\; ,
\end{eqnarray}
with the curvature  scalar
\begin{equation} 
R(t)=6\left[\dot{H}(t)+2H^2(t)\right]\; ,
\end{equation}
and the Hubble expansion rate 
\begin{equation} 
H(t)=\frac{\dot{a}(t)}{a(t)}\; .
\end{equation}

\section{Nonequilibrium equations for the scalar field}
The Lagrangian density of a $\phi^4$-theory in curved space-time
is given by
\begin{equation} 
{\cal L}=\sqrt{-g}\left\{\textstyle{\frac 1 2} \partial_\mu \Phi \partial^\mu \Phi
-\textstyle{\frac 1 2} m^2 \Phi^2
-\frac \xi 2  R \Phi^2 -\frac{\lambda}{4!} \Phi^4\right\}
\; ,\end{equation}
where $R(x)$ is the curvature scalar and $\xi$ the
 bare dimensionless parameter 
describing the coupling of the bare scalar field to the gravitational
background. 
We split the field $\Phi$ into its expectation value $\phi$
and the quantum fluctuations $\psi$:
\begin{equation} 
\Phi(\vec x,t)=\phi(t)+\psi(\vec x,t)\; ,
\end{equation}
with
\begin{equation} 
\phi(t)=\langle\Phi(\vec x,t)
\rangle=\frac{{\rm Tr}\Phi\rho(t)}{{\rm Tr}\rho(t)}
\; ,
\end{equation}
where $\rho(t)$ is the density matrix of the system which satisfies the 
Liouville equation
\begin{equation} 
i\frac{d\rho(t)}{dt}=[{\cal H}(t),\rho(t)]
\; .\end{equation}
The one-loop equation of motion of a scalar field with
$\lambda \phi^4$ interaction
has been obtained in the FRW universe by Ringwald \cite{Ringwald:1988};
we follow closely his formulation. 
The equation of motion for the classical field is
\begin{equation} 
\ddot{\phi}+ 3 H\dot{\phi}+(m^2+\xi R)\phi+\frac{\lambda}{6}\phi^3+
\frac{\lambda}{2}
\langle\psi^2\rangle\phi=0
\; .\end{equation}
The expectation value of the quantum fluctuations 
$\langle\psi^2\rangle$ can be expressed as
\begin{equation} 
\langle\psi^2\rangle=-i G(t,\vec x;t,\vec x)
\end{equation}
in terms of the
non-equilibrium Green function $G(t,\vec x;t',\vec x')$ which 
satisfies 
\begin{equation}
\label{comtmode}
\left[\frac{{\partial}^2}{{\partial}t^2}+3H\frac{{\partial}}
{{\partial}t}+ a^{-2}(t)\vec \nabla^2+m^2+\xi R(t)+
\frac{\lambda}{2}\phi^2(t)\right]G(t,\vec x;t',\vec x')=
\frac{i}{a^3(t)}\delta(t,\vec x;t',\vec x')\; .
\end{equation}
The boundary conditions for this Green functions will be 
given below.
Due to the presence of the term $H(t)\partial/\partial t$
the differential operator on the left hand side of this
equation is  non-hermitian.
It is made hermitian by introducing conformal time 
and appropriate scale factors. Conformal time is defined as
\begin{equation} 
\tau=\int\limits_{0}^{t}\!dt'\frac{1}{a(t')} 
\; .\end{equation}
In conformal time the line-element (\ref{line}) reads
\begin{equation} 
\label{newline}
{d}s^2=C(\tau)({d}\tau^2-{d}{\vec x} ^2)
\; ,\end{equation}
where the conformal factor $C(\tau)$ is given by
\begin{equation} 
C(\tau)=a^2(\tau)
\; .\end{equation}
We further rescale the scalar field and its quantum fluctuations 
 by introducing the dimensionless `conformal' fields
\begin{eqnarray}
\varphi(\tau)&=& a(t)\phi(t) \; , \\
\tilde{\psi}(\tau,\vec x)&=& a(t) \psi(t,\vec x) \; .
\end{eqnarray}
The Green function is rescaled accordingly via
\begin{equation} 
\tilde{G}(\vec x,\tau;\vec x ',\tau')=a(t)a(t ')G(\vec x
,t;\vec x,t')\; .
\end{equation}
The equation of motion of the classical field $\varphi(\tau)$
now becomes
\begin{equation} 
\varphi'' + a^2\left[m^2+\left(\xi-{\textstyle \frac{1}{6}}\right)R\right]\varphi
+\frac \lambda 6 \varphi^3 + \frac{\lambda}{2i}\varphi
\tilde{G}(\tau,\vec x;\tau,\vec x)=0\; ,
\end{equation}
where the primes denote derivatives with respect to conformal time.
The two-point-function $\tilde{G}$ now satisfies
\begin{equation} 
\left[ \frac{\partial^2}{\partial \tau^2}-\nabla^2+M^2(\tau)\right]
\tilde{G}(\vec x
,\tau;\vec x,\tau ')=-\delta(\vec x, \tau;\vec x ',\tau ')
\; .
\end{equation}
Here $M^2(\tau)$ denotes the square of the effective mass term
the fluctuation field $\tilde \psi(\tau,\vec x)$
\begin{equation} 
\label{effm}
M^2(\tau)=C(\tau)\left[ m^2+\left(\xi-{\textstyle \frac{1}{6}} \right)R(\tau)+\frac
\lambda 2 \frac{\varphi^2(\tau)}{a^2(\tau)}
\right]\; .
\end{equation}
The problem of determining the Green function is now essentially
reduced to the equivalent problem in Minkowski space.
We expand the fluctuation field in terms of the mode functions
$U_k(\tau)\exp(i\vec k \vec x)$ via
\begin{equation} 
\tilde{\psi}(\tau,\vec x) =
\int\! \frac{d^3k}{(2\pi)^3} \left[c(\vec k) U_k(\tau)\exp(i\vec k \vec x)
+c^\dagger(\vec k)U^*_k(\tau)\exp(-i\vec k \vec x)\right]
\; .\end{equation}
The functions $U_k(\tau)$ satisfy the mode equation
\begin{equation} 
U_k''(\tau)+\Omega^2_k(\tau)U_k(\tau)=0\; ,
\end{equation}
with
\begin{equation} 
\Omega_k^2(\tau)=k^2+a^2(\tau)\left[
m^2+\left(\xi - {\textstyle \frac{1}{6}}\right)R(\tau)\right]+\frac \lambda 2
\varphi^2(\tau)
\; .\end{equation}
We further impose  the initial conditions
\begin{eqnarray}
\label{incon}
U_k(0)=1&\mbox{ ; }&U'_k(0)=-i\Omega_k(0)
\; ,
\end{eqnarray}
with
\begin{equation} 
\Omega_k(0)=\sqrt{k^2+M^2(0)}
\; .
\end{equation}
In the following we will use the 
short notation $\Omega_{k0}=\Omega_k(0)$.
The nonequilibrium Green function $\tilde{G}_k(\tau,\vec x;\tau '
,\vec x')$ can by expressed in terms of the mode functions
via
\begin{eqnarray}
\tilde{G}_k(\tau,\vec x;\tau '
,\vec x')&=&\int\! \frac{d^3k}{(2\pi)^3}\frac{i}{2\Omega_{k0}}
\left\{\theta(\tau-\tau ')U_k(\tau)U_k^{*}(\tau ')
\exp(i\vec k(\vec x-\vec x')) \right.\nonumber \\
&&\hspace{2.5cm}\left.+\theta(\tau ' - \tau)U_k^{*}(\tau ')U_k(\tau)
\exp(-i\vec k(\vec x-\vec x'))\right\}
\; .
\end{eqnarray}
The expectation value of the fluctuation fields is given,
therefore, by the fluctuation integral
\begin{equation} \label{flucdef}
{\cal F}(\tau) =\langle \tilde{\psi}^2(\tau)\rangle=
-i\tilde G (\tau,\vec x;\tau,\vec x)=
\int\! \frac{d^3k}{(2\pi)^3}\frac{|U_k(\tau)|^2}{2\Omega_{k0}}
\; .
\end{equation}
The unrenormalized equation of motion of the inflaton field
reads
\begin{equation} 
\varphi'' + a^2(\tau)\left[m^2+\left(\xi-{\textstyle \frac{1}{6}}\right)R(\tau)\right]
\varphi
+\frac \lambda 6 \varphi^3 + \frac{\lambda}{2}\varphi {\cal F}(\tau)
=0 \; . \end{equation}
The regularization of the fluctuation integral and the renormalized
form of this equation will be discussed below.

The boundary conditions for the mode functions and the corresponding
definition of the fluctuation integral are related to the definition
of the initial quantum state and/or density matrix 
of our nonequilibrium expansion. It would, e.g., look different
if we had used the time variable $t$ instead
of conformal time $\tau$. These initial conditions, 
a conformal vacuum state \cite{Birrell:1982},
 have already been chosen in \cite{Ringwald:1987a},
they are discussed to some extent in \cite{Boyanovsky:1997a}.
The initial conditions for the cosmological expansion are
even more subtle since even in principle we cannot start from an 
equilibrium situation, except if the classical field is at 
a minimum of the effective potential and the energy momentum
tensor vanishes there -  obviously not an interesting situation.
When considering nonequilibrium processes
in Minkowski space-time one can imagine that an initial classical field
is maintained away from the minimum of the effective potential 
by a source (like capacitor plates for an electric field) which
is switched off at time zero. In such a situation it makes sense
to assume that
the initial state is the vacuum state corresponding to that initial
classical field. In the cosmological context the presence of a 
nonvanishing classical energy leads already to a cosmological
expansion which in turn influences the quantum fluctuations.
So starting with the conformal vacuum state  or a thermal
state constructed on it seems a priori artificial. 
Since there is no solution of principle to this problem  
one has to hope (or try out by numerical experiments)
that the precise initial conditions do not influence too
strongly that period of the cosmological expansion one wants
to study. 
\section{The energy momentum tensor}
In order to formulate Einstein's field equation we have to
discuss the energy momentum tensor of the scalar field in
curved space time. For a classical field it reads \cite{Birrell:1982}
\begin{eqnarray}
T_{\mu\nu}&=&(1-2\xi)\phi_{;\mu}\phi_{;\nu}+\left(2\xi-\textstyle{\frac 1 2}\right)
g_{\mu\nu}g^{\rho\sigma}\phi_{;\rho}\phi_{;\sigma}-
2\xi\phi_{;\mu\nu}\phi\nonumber\\
&&+2\xi g_{\mu\nu}\phi\Box\phi-\xi G_{\mu\nu}\phi^2
+\textstyle{\frac 1 2} m^2g_{\mu\nu}\phi^2
+\frac{\lambda}{4!}g_{\mu\nu}\phi^4\; .
\end{eqnarray}
In the conformally flat FRW metric the energy momentum tensor
is diagonal. One obtains 
for its time-time component and its trace 
\begin{eqnarray}
T_{tt}^{cl}&=&\textstyle{\frac 1 2}\dot{\phi}^2
+\textstyle{\frac 1 2} m^2\phi^2+\frac{\lambda}{4!}\phi^4
-\xi G_{tt}\phi^2+6\xi H\phi\dot{\phi}\; ,\nonumber\\
T_{\mu}^{\mu\;cl}&=&-\dot{\phi}^2
+2m^2\phi^2+\frac{\lambda}{6}\phi^4
-\xi G_{\mu}^{\mu}\phi^2
+6\xi(\phi\ddot{\phi}+\dot{\phi}^2+3H\phi\dot{\phi})\; . 
\end{eqnarray}
We introduce again conformal time and the conformal rescaling of the
fields. Furthermore, we include the quantum fluctuations of the field
$\varphi$.
We obtain \footnote{We continue to consider $T_{tt}$ instead of
$T_{\tau\tau}=a^2T_{tt}$ for convenience.}
\begin{eqnarray}
{T_{tt}}&=&\frac{1}{2a^4}\varphi'^2+\frac{1}{2a^2} m^2\varphi^2
+\frac{\lambda}{4!a^4}\varphi^4
+(1-6\xi)\left(
\frac{H^2}{2a^2}\varphi^2-\frac{H}{a^3}\varphi\varphi'
\right)
\nonumber\\
&&+\frac{1}{a^2}\int\! \frac{d^3k}{(2\pi)^3}\frac{1}{2\Omega_{k0}^2}\Biggl\{
\frac{1}{2a^2}|U_k'|^2+\frac{1}{2a^2}\Omega(\tau)|U_k|^2
-{\textstyle{\textstyle\frac{1}{2}}}\left(\xi -{\textstyle \frac{1}{6}}\right)R|U_k|^2\nonumber\\
&&\hspace{3.6cm}-{\textstyle {\textstyle\frac{1}{2}} }(6\xi-1) H^2|U_k|^2+{\textstyle {\textstyle\frac{1}{2}} }
(6\xi -1)\frac{H}{a}\frac{d}{d\tau}|U_k|^2
\Biggr\}\; ,\nonumber\\
\end{eqnarray}
and
\begin{eqnarray}
T_\mu^\mu&=&(1-6\xi)\left(-\frac{\varphi'^2}{a^4}
-\frac{H^2}{a^2}\varphi^2
+2\frac{H}{a^3}\varphi\varphi'\right) +6\frac{\xi}{a^4}\varphi\varphi''
+2m^2\frac{\varphi^2}{a^2}+\frac{\lambda}{6a^4}\varphi^4
\nonumber\\
&&+\frac{1}{a^2}\int\! \frac{d^3k}{(2\pi)^3}\frac{1}{2\Omega_{k0}}\left\{
(1-6\xi)\left[\left(\frac{|U_k'|^2}{a^2}
+\Omega^2(\tau)\frac{|U_k|^2}{a^2}
\right)\right.\right.\nonumber\\
&& \left.\hspace{5cm}-\left(
2\frac{|U_k'|^2}{a^2}+H^2|U_k|^2-\frac{H}{a}\frac{d}{d\tau}|U_k|^2
\right)\nonumber\right]\\
&&\left.\hspace{3cm}+\left[m^2+
\frac{\lambda}{2}\frac{\varphi^2}{a^2}-\left(\xi-{\textstyle \frac{1}{6}}\right)  R
\right]|U_k|^2\right\}\; .
\end{eqnarray}
Energy density and pressure are
related to the energy momentum tensor via
\begin{eqnarray}
T_{tt} &=& {\cal E} \nonumber\; , \\
T_\mu^\mu&=& {\cal E} -3 p\; .
\end{eqnarray}   
It is straightforward to show, using the equations of motion
for the classical field and for the mode functions
 (\ref{comtmode}),
that the energy is covariantly conserved:
\begin{equation} 
{\cal E}'(\tau)/a(\tau)+3H(\tau)(p(\tau)+{\cal E}(\tau))=0\; .
\end{equation}
\section{Perturbative expansion}
In order to prepare the renormalized version of the equations given
in the previous section we introduce a suitable expansion of the
mode functions, which was successfully used
in \cite{Baacke:1997a,Baacke:1997b} for the inflaton field
coupled to itself, to fermions, and  gauge bosons in Minkowski-space.	
Adding the term $M^2(0)$
on both sides of the mode function equation it takes 
the form
\begin{equation} 
\label{udgl}
\left[ \frac{d^2}{d\tau^2}+
\Omega_{k0}^2\right]U_k(\tau)=-V(\tau)U_k(\tau)\; ,
\end{equation}
with
\begin{eqnarray}
V(\tau)&=&M^2(\tau)-M^2(0)\; ,
\nonumber\\
\Omega_{k0}&=&\left[\vec k^2+M^2(0)\right]^{1/2}
\end{eqnarray}
(for the definition of $M^2(\tau)$ see eq.(\ref{effm})).
The first and second derivatives of this potential 
will be needed in the perturbative
expansion and are given by
\begin{eqnarray}
\label{vfirder}
V'(\tau)&=&2aHM^2(\tau)+a^2\left(\xi-{\textstyle \frac{1}{6}}\right)R'+\lambda \varphi
 \varphi'-\lambda a H\varphi^2 \; ,
\\
\label{vsecder}
V''&=&{\textstyle {\textstyle \frac{1}{3}}} a^2 R M^2-2 a^2 H^2 M^2+
2a HV'(\tau)\nonumber\\
&&+2 a^3H\left(\xi-{\textstyle \frac{1}{6}} \right)R'-2\lambda a H \varphi\varphi'
+a^2\left(\xi-{\textstyle \frac{1}{6}}\right)R''
\nonumber\\
&&+\lambda(\varphi\varphi''+\varphi'^2)
-\lambda\frac{a^2}{6}R\varphi^2 + \lambda a^2 H^2\varphi^2\; .
\end{eqnarray}
 Including
 the initial conditions (\ref{incon}) 
the mode functions satisfy the equivalent integral equation
\begin{equation} 
U_k(\tau)=e^{-i\Omega_{k0} \tau}+
\int\limits^{\infty}_{0}\!{d}\tau'
\Delta_{k,{\rm ret}}(\tau-\tau')V(\tau')U_k(\tau')\;,
\end{equation}
with
\begin{equation} 
\label{fvt}
\Delta_{k,{\rm ret}}(\tau-\tau')= -\frac{1}{\Omega_{k0}}
\Theta(\tau-\tau')\sin\left(\Omega_{k0}(\tau-\tau')\right) \; .
\end{equation}
We separate $U_k(\tau)$ into the trivial part corresponding to
the case $V(\tau)=0$ and a function $h_k(\tau)$ which represents the
reaction
to the potential by making the ansatz
\begin{equation} 
\label{ansatz}
U_k(\tau)=e^{-i\Omega_{k0} \tau}(1+h_k(\tau)) \; .
\end{equation}
$h_k(\tau)$ satisfies then the integral equation
\begin{equation} \label{finteq}
h_k(\tau)=\int\limits^{\tau}_{0}\!{d}\tau'\Delta_{k,{\rm ret}}
(\tau-\tau')V(\tau')(1+h_k(\tau'))e^{i\Omega_{k0} (\tau-\tau')}\;,
\end{equation}
and an equivalent differential equation
\begin{equation} \label{fdiffeq}
h_k''(\tau)-2i\Omega_{k0}h'_k(\tau)=-V(\tau)(1+h_k(\tau))\;,
\end{equation}
with the initial conditions $h_k(0)=\dot{h}_k(0)=0$.
We expand now $h_k(\tau)$ with respect to orders in $V(\tau)$
by writing
\begin{eqnarray}
\label{entwicklung}
h_k(\tau)&=& h_k^{(1)}(\tau)+h_k^{(2)}(\tau)+h_k^{(3)}(\tau) +\cdots \\
 &=& h_k^{(1)}(\tau)+h_k^{{(\overline{2})}}(\tau)
\; ,\end{eqnarray}
where $h_k^{(n)}(\tau)$ is of n'th order in $V(\tau)$ and 
$h_k^{{{(\overline{n})}}}(\tau)$
is the sum over all orders beginning with the n'th one:
\begin{equation} 
h_k^{(\overline{n})}(\tau)=\sum_{l=n}^\infty h_k^{(n)}(\tau)
\; .\end{equation}
The $h_k^{(n)}$ are obtained by iterating the integral
equation (\ref{finteq}) or the differential equation
(\ref{fdiffeq}). The function $h_k^{{(\overline{1})}}(\tau)$ is
identical to the function $h_k(\tau)$ itself which is obtained
by solving (\ref{fdiffeq}). The function
$h_k^{{(\overline{2})}}(\tau)$ can again be obtained
by iteration via
\begin{equation} \label{f2inteq}
h_k^{{(\overline{2})}}(\tau)=
\int\limits^{\tau}_{0}\!{d}\tau'\Delta_{k,{\rm ret}}
(\tau-\tau')V(\tau')
h_k^{{(\overline{1})}}(\tau')e^{i\Omega_{k0} (\tau-\tau')} \;,
\end{equation}
or, using the differential equation,
via
\begin{equation} \label{f2diffeq}
{h''}_k^{{(\overline{2})}} (\tau)-2i\Omega_{k0}
{h'}_k^{{(\overline{2})}}(\tau)=-V(\tau)h_k^{{(\overline{1})}}(\tau) \; .
\end{equation}
This iteration has the numerical aspect that it avoids computing
$h_k^{{(\overline{2})}}$ via the small difference
$h_k^{{(\overline{1})}}-h_k^{(1)}$. However, the integral equations
are used as well in order to derive the asymptotic behaviour as
$\Omega_{k0}\to \infty$ and to separate divergent and finite 
contributions.
We will give here the relevant leading terms for $h_k^{(1)}(\tau)$ and
$h_k^{(2)}(\tau)$. We have
\begin{equation} 
h_k^{(1)}(\tau)=\frac{i}{2 \Omega_{k0}}
\int\limits^{\tau}_{0}\!{d}\tau'
(\exp(2 i \Omega_{k0}(\tau-\tau'))-1)V(\tau')  \; .
\end{equation}
Integrating by parts we obtain
\begin{equation} \label{f1exp}
h_k^{(1)}(\tau)=
-\frac{i}{2\Omega_{k0}}\int\limits^{\tau}_{0}\!{d}\tau'
V(\tau')-\frac{1}{4\Omega_{k0}^2}V(\tau)
+\frac{1}{4\Omega_{k0}^2}\int\limits^{\tau}_{0}\!{d}\tau'
\exp(2 i \Omega_{k0}(\tau-\tau'))V'(\tau')\;,
\end{equation}
or, by another integration by parts,
\begin{eqnarray}
h_k^{(1)}(\tau)&=&
-\frac{i}{2\Omega_{k0}}\int\limits^{\tau}_{0}\!{d}\tau'
V(\tau')-\frac{1}
{4\Omega_{k0}^2}V(\tau)+\frac{i}{8\Omega_{k0}^3} V'(\tau) \\
&&-\frac{i}{8\Omega_{k0}^3}\int\limits^{\tau}_{0}\!{d}\tau'
\exp(2 i \Omega_{k0} (\tau-\tau'))V''(\tau') \; .
\end{eqnarray}
We will need often the real part of $h^{(1)}_k$ for which we find
\begin{eqnarray}  \label{realexp}
 {\rm Re}\;h^{(1)}_k(\tau) &=&-\frac{1}{4\Omega_{k0}^2}V(\tau)
+ \frac{1}{4\Omega_{k0}^2}\int\limits^{\tau}_{0}\!{d}\tau'
\cos(2  \Omega_{k0}(\tau-\tau'))V'(\tau')\nonumber \\
&=&-\frac{1}{4\Omega_{k0}^2}V(\tau)
+ \frac{1}{4\Omega_{k0}^2}{\cal C}(V',\tau)\; .
\end{eqnarray}
Since various Fourier integrals will appear in the finite parts
of the fluctuation integrals we introduce for later
convenience the notations
\begin{eqnarray}
{\cal S}(f,\tau)&=&\int\limits^{\tau}_{0}\!{d}\tau'
\sin(2  \Omega_{k0}(\tau-\tau'))f(\tau') \; ,\\
{\cal C}(f,\tau)&=&\int\limits^{\tau}_{0}\!{d}\tau'
\cos(2  \Omega_{k0}(\tau-\tau'))f(\tau') \; .
\end{eqnarray}
For the leading behaviour of $h_k^{(2)}(\tau)$ we find
\begin{equation} \label{f2exp}
h_k^{(2)}(\tau)= -\frac{1}{4\Omega_{k0}^2}
\int\limits^{\tau}_{0}\!{d}\tau'\int\limits^{t'}_{0}\!{d}\tau''
V(\tau')V(\tau '') + O(\Omega_{k0}^{-3}) \; .
\end{equation}
In terms of this perturbative expansion we can write the mode functions
appearing in the fluctuation integrals in the equation of motion and in
the energy momentum tensor as
\begin{equation} 
|U_k|^2=1+2 {\rm Re }\, h_k^{(\overline{1})}+|{h_k^{(\overline{1})}}|^2
\; ,\end{equation}
and
\begin{eqnarray}
|U'_k|^2&=&
\Omega_{k0}^2\left(1+2{\rm Re }\, h_k^{(\overline{1})}+|{h_k^{(\overline{1})}}|^2\right)
+|h_k^{\prime(\overline{1})}|^2\nonumber\\
&&\hspace{1cm}-i\Omega_{k0}\left(-2i{\rm Im }\,{h'_k}^{(\overline{1})}
-2i{\rm Im }\, h_k^{(\overline{1})*}{h}_k^{\prime(
\overline{1})}\right)
\; .\end{eqnarray}
As the potential is real, the leading behaviour of the sums
is
\begin{eqnarray}
\label{reskalsum}
1+2{\rm Re }\,{h}_k^{(\overline{1})}+
|h_k^{(\overline{1})}|^2&=&1-\frac{1}{2\Omega_{k0}^2}V(\tau)+
\frac{1}{4\Omega_{k0}^3}\sin(2\Omega_{k0}\tau)V'(0)+
\frac{1}{8\Omega_{k0}^4}V''(\tau)\nonumber\\
&&-\frac{1}{8\Omega_{k0}^4}
(2\Omega_{k0}\tau)V''(0)
+\frac{3}{8\Omega_{k0}^4}V^2(\tau)+{\cal O}(\Omega_{k0}^{-5})\;,
\end{eqnarray}
and
\begin{eqnarray}
\label{imskalsum}
-2i{\rm Im }\,{h}_k^{\prime(\overline{1})}-2i{\rm Im }\, h_k^{(\overline{1})*}{h}_k^
{\prime(\overline{1})}&=&\frac{i}{\Omega_{k0}}V(\tau)
-\frac{i}{2\Omega_{k0}^2}\sin(2\Omega_{k0}\tau)V'(0)
-\frac{i}{4\Omega_{k0}^3}V''(\tau)\nonumber\\
&&+\frac{i}{4\Omega_{k0}^3}\cos(2\Omega_{k0}\tau)V''(0)
-\frac{3i}{4\Omega_{k0}^3}V^2(\tau)+{\cal
O}(\Omega_{k0}^{-4})\; .\nonumber\\
\end{eqnarray}
From the Wronskian relation
\begin{equation} 
U_k{U_k^*}'-U_k'U_k^*=2i\Omega_{k0}
\end{equation}
we obtain the relation
\begin{equation} 
2i\Omega_{k0}\left(2{\rm Re }\,{h}_k^{(\overline{1})}+
|h_k^{(\overline{1})}|^2\right)
-2i{\rm Im }\,{h}_k^{\prime
(\overline{1})}-2i{\rm Im }\, h_k^{(\overline{1})*}{h}_k^{\prime(
\overline{1})}=0
\; ,
\end{equation}
which proves to be useful in simplifying the mode integrals
occuring in the energy momentum tensor.

\section{Renormalization of the equation of motion}
The expansion of the mode functions 
allows  a free choice of the regularization
scheme. We  will use dimensional regularization here since
in the presence of quadratic and quartic divergencies it 
is much easier to handle than in Pauli-Villars regularization.
We will show that in this scheme the divergence structure 
of the equation of motion and of the energy 
momentum tensor  have 
the correct form for a consistent renormalization.
Furthermore, we will present
 the explicit form of the finite parts of the
fluctuation integrals occuring in
the equation of motion and the energy momentum tensor.

The fluctuation integral of the equation of motion can be split into a 
divergent and a convergent part.
Using (\ref{realexp}) we obtain
\begin{eqnarray}
\cal F(\tau)&=&\int\! \frac{d^3k}{(2\pi)^3}\frac{1+2{\rm Re}\,h_k^{(\overline{1})}(\tau)+
|{h_k^{(\overline{1})}}(\tau)|^2}{2\Omega_{k0}}\nonumber\\
&=&\int\! \frac{d^3k}{(2\pi)^3}\frac{1}{2\Omega_{k0}}\left(1-\frac{1}{2{\Omega_{k0}}^2}
V(\tau)+\frac{1}{2{\Omega_{k0}}^2}{\cal C}(V',\tau)\right.\nonumber\\
&&\left.\hspace{2.7cm}
+2{\rm Re }\,{h}_k^{(\overline{2})}(\tau)+|h_k^{(\overline{1})}(\tau)|^2\right)
\; .
\end{eqnarray}
The first two terms in the integrand  have to be regularized. 
We first rewrite
the basic equation of motion, including appropriate counter terms, as
\begin{equation} 
\varphi ''+ a^2\left[m^2+\delta m+\left(\xi-{\textstyle \frac{1}{6}} +
\delta \xi\right) R\right]\varphi
+\frac {\lambda+\delta \lambda}{6} \varphi^3 + \frac{\lambda}{2}\varphi
{\cal F}=0\;.
\end{equation}
Next we separate from the term $\lambda \varphi {\cal F}/2$  the
dimensionally regularized divergent parts
\begin{eqnarray}
\label{dimreg1}
\left\{\frac{\lambda}{2}\varphi(\tau)
\int\!\frac{{d^3}k}{(2\pi)^3}\, \frac{1}{2\Omega_{k0}}\right\}_
{{\rm reg}}
&=&\mu^\epsilon\frac{\lambda}{2}\varphi(\tau)\int
\!\frac{{d^d} k}{(2\pi)^d}\,
\frac{1}{2\left[\vec k^2+M^2(0)\right]^{1/2}}
\nonumber\\
&=&-\frac{\lambda M^2(0) 
\varphi(\tau)}{32\pi^2}\left\{\frac{2}{\epsilon}+
\ln{\frac{4\pi\mu^2}{M^2(0)}}-\gamma+1\right\}\;,
\end{eqnarray}
and
\begin{eqnarray}
\label{dimreg2}
\left\{-\frac{\lambda}{2}\varphi(\tau)
V(\tau)\int\!\frac{{d^3}k}{(2\pi)^3}\, \frac{1}{4{\Omega_{k0}}^3}
\right\}_{{\rm reg}}&=&-\frac{1}{8}(\mu^2)
^\epsilon\lambda\varphi(\tau) V(\tau)\int\!
\frac{{d^d} k}{(2\pi)^d}\,
\frac{1}{\left[\vec k^2+M^2(0)\right]^{3/2}}
\nonumber\\
&=& -\frac{\lambda \varphi(\tau)V(\tau)}
{32\pi^2}\left\{\frac{2}{\epsilon}
+\ln{\frac{4\pi\mu^2}{M^2(0)}}-\gamma\right\}\; .
\end{eqnarray}
Recalling that $V(\tau)=M^2(\tau)-M^2(0)$ and 
$M^2(\tau)=a^2(\tau)\left[m^2+(\xi-1/6)R\right]+\lambda\varphi^2/2$
one sees that $M^2(0)$ cancels for the divergent terms and 
that therefore
the counter terms can be chosen independent of the initial conditions
as
\begin{eqnarray}                 \label{deltam}
\delta m^2 &=&
\frac{\lambda m^2}{32\pi^2}\left\{\frac{2}{\epsilon}+
\ln{\frac{4\pi\mu^2}{m^2}}-\gamma\right\} \; ,\\
\label{deltalam}
\delta \lambda &=&
\frac{3\lambda^2}{32\pi^2}\left\{\frac{2}{\epsilon}+
\ln{\frac{4\pi\mu^2}{m^2}}-\gamma\right\}\; , \\
\delta \xi& =&
\frac{\lambda(\xi-{\textstyle \frac{1}{6}})}{32\pi^2}\left\{\frac{2}{\epsilon}+
\ln{\frac{4\pi\mu^2}{m^2}}-\gamma\right\}\; .
\end{eqnarray}
The renormalized equation of motion now reads
\begin{equation} 
\varphi '' + a^2\left[m^2+\Delta m+(\xi+
\Delta \xi-\frac 1 6) R\right]\varphi
+\frac {\lambda+\Delta \lambda}{6} \varphi^3
+ \frac{\lambda}{2}\varphi
{\cal F}_{fin}=0\;,
\end{equation}
with
\begin{equation} 
{\cal F}_{fin}=-\frac{M^2(0)}{16\pi^2}+\int\! \frac{d^3k}{(2\pi)^3}\frac{1}{2\Omega_{k0}}
\left(
\frac{1}{2{\Omega_{k0}}^2}{\cal C}(V',\tau)
+2{\rm Re }\,{h}_k^{(\overline{2})}+|h_k^{(\overline{1})}|^2\right)\; ,
\end{equation}
and with the finite corrections
\begin{eqnarray}
\Delta m^2 &=&-\frac{\lambda m^2}{32\pi^2}\ln\frac{m^2}{M^2(0)}\; ,
\\
\Delta \lambda&=&-\frac{3\lambda^2}{32\pi^2}\ln\frac{m^2}{M^2(0)}\; ,
\\
\Delta \xi& =&-\frac{\lambda(\xi-{\textstyle \frac{1}{6}})}
{32\pi^2}\ln\frac{m^2}{M^2(0)} \; .
\end{eqnarray}

\section{Renormalization of the energy momentum tensor}
In order to derive a renormalized form of the Friedmann equations
we have to renormalize the energy momentum tensor as well. 
In principle this 
has been discussed long ago and it will not be a surprise that
the divergent parts are in one-to-one correspondence to those
given e.g. in \cite{Birrell:1982}. However, we have
to discuss this subject in the framework of
nonequilibrium quantum field theory,
and we are interested in particular in the precise form of the 
finite parts which will be the subject
of a numerical computation.  In order to renormalize
the energy we introduce the available counter terms
into the unrenormalized expression so that it reads now
\begin{eqnarray} \label{Ewct}
{\cal E}&=&
\frac{1}{2a^4}\varphi'^2+\frac{1}{2a^2} (m^2+\delta m^2)\varphi^2
+\frac{\lambda+\delta \lambda}{4!a^4}\varphi^4\nonumber\\
&&-6\left(\xi-{\textstyle \frac{1}{6}}+\delta \xi\right)\left(
\frac{H^2}{2a^2}\varphi^2-\frac{H}{a^3}\varphi\varphi'
\right)+\delta \tilde{\Lambda}+\delta \tilde{\alpha} H_{tt}
+\delta\tilde{Z}G_{tt}\nonumber\\
&&+\frac{1}{a^2}\int\! \frac{d^3k}{(2\pi)^3}
\left\{\frac{\Omega_{k0}}{2a^2}
\left(1+2{\rm Re }\,{h}_k^{(\overline{1})}+|h_k^{(\overline{1})}|^2
\right)+\frac{1}{4\Omega_{k0}a^2}
|{h}_k^{\prime(\overline{1})}|^2\right.\nonumber\\
&&\hspace{3cm}-\frac{i}{4a^2}\left(-2i{\rm Im }\,{h}_k^{\prime(\overline{1})}-2i
{\rm Im }\, h_k^{(\overline{1})*}h_k^{\prime(
\overline{1})}\right)
\nonumber\\
&&\hspace{3cm}+\frac{1}{4\Omega_{k0}a^2}V(\tau)
\left(1+2{\rm Re }\,{h}_k^{(\overline{1})}+|h_k^{(\overline{1})}|^2\right)\nonumber\\
&&\hspace{3cm}-\frac{1}{4\Omega_{k0}}(6\xi-1)\left(\frac R 6+H^2\right)
\left(1+2{\rm Re }\,{h}_k^{(\overline{1})}+|h_k^{(\overline{1})}|^2\right)
\nonumber\\
&&\hspace{3cm}\left.+
\frac{1}{4\Omega_{k0}} (6\xi-1)\frac H a \frac{d}{d\tau}
\left(1+2{\rm Re }\,{h}_k^{(\overline{1})}+|h_k^{(\overline{1})}|^2\right)\right\}
\; .
\end{eqnarray}
The divergent parts of the fluctuation integral are 
\begin{eqnarray}
{\cal E}_{div,fluc}&=&
\frac{1}{a^2}\int\! \frac{d^3k}{(2\pi)^3}\left[\frac{\Omega_{k0}}{2a^2}
+\frac{1}{4\Omega_{k0}a^2}V(\tau)-\frac{1}{16\Omega_{k0}^3a^2}V^2(\tau)
\right.\nonumber\\
&&\hspace{2.3cm}-{\textstyle\frac{1}{2}}(6\xi-1)\left(\frac R 6 +H^2\right)
(\frac{1}{2\Omega_{k0}}-\frac{1}{4\Omega_{k0}^3}V(\tau))\nonumber\\
&&\hspace{2.3cm}\left.-{\textstyle\frac{1}{2}}(6\xi-1)\frac{H}{a}\frac{1}
{4\Omega_{k0}^3}V'(\tau)\right]\; .
\end{eqnarray}
Dimensional regularisation of the divergent integrals yields
\begin{eqnarray}\label{fdiv}
&&\int\! \frac{d^3k}{(2\pi)^3}\frac{1}{a^2}\left[\frac{\Omega_{k0}}{2}
+\frac{1}{4\Omega_{k0}}V(\tau)-
\frac{1}{16\Omega_{k0}^3}V^2(\tau)\right]
\nonumber\\
&&=-\frac{M^4(\tau)}{64\pi^2a^2}\left\{\frac{2}{\epsilon}
+\ln{\frac{4\pi\mu^2}{M^2(0)}}-\gamma\right\}
+\frac{M^4(0)}{128\pi^2a^2}-\frac{M^2(0)M^2(\tau)}{32\pi^2a^2}
\\
&&=-\frac{a^2\left[m^2+\left(\xi-{\textstyle \frac{1}{6}}\right)R 
+\frac{\lambda}{2}\varphi^2\right]^2}{64\pi^2}\left\{\frac{2}{\epsilon}
+\ln{\frac{4\pi\mu^2}{M^2(0)}}-\gamma\right\}
+\frac{M^4(0)}{128\pi^2a^2}-\frac{M^2(0)M^2(\tau)}{32\pi^2a^2}\nonumber
\; .
\end{eqnarray}
The terms proportional $\lambda m^2 \varphi^2$ and
$\lambda^2/4 \varphi^4$ in (\ref{fdiv}) are cancelled by the mass
and coupling constant counter terms. The divergent term which depends on
$m^4$ determines the cosmological constant
counter term, that is
\begin{equation} 
\delta \tilde{\Lambda} =\frac{m^4}{64\pi^2}\left\{\frac{2}{\epsilon}
+\ln{\frac{4\pi\mu^2}{m^2}}-\gamma\right\}\; .
\end{equation}
The remaining terms in (\ref{fdiv}) combine with
the corresponding expressions
of the following two dimensionally regularized integrals
\begin{eqnarray}\label{sdiv}
&&-{\textstyle\frac{1}{2}}\left(6\xi-1\right)
(\frac R 6 +H^2)\int\! \frac{d^3k}{(2\pi)^3}\left[\frac{1}{2\Omega_{k0}}
-\frac{1}{4\Omega_{k0}^3}V(\tau)
\right]\nonumber\\
&&=(6\xi-1)\left(\frac R 6 +H^2\right)
\frac{M^2(\tau)}{32\pi^2}\left\{\frac{2}{\epsilon}
+\ln{\frac{4\pi\mu^2}{M^2(0)}}-\gamma\right\}\nonumber\\
&&\hspace{1cm}+\frac{1}{32\pi^2}
(6\xi-1)\left(\frac R 6+H^2\right)M^2(0)\;,
\end{eqnarray}
and
\begin{eqnarray}\label{tdiv}
&&-{\textstyle\frac{1}{2}}(6\xi-1)\frac{H}{a}\int\! \frac{d^3k}{(2\pi)^3}
\frac{1}{4\Omega_{k0}^3}V'(\tau)
\nonumber\\
&&=-\frac{1}{16\pi^2}(6\xi-1)H^2M^2(\tau)\left\{\frac{2}{\epsilon}
+\ln{\frac{4\pi\mu^2}{M^2(0)}}-\gamma\right\}\nonumber\\
&&\hspace{0.5cm}-\frac{1}{32\pi^2}(6\xi-1)\frac H a
\left[\left(\xi-{\textstyle{\frac1 6}}\right)R'a^2+\lambda \varphi\varphi'
-\lambda a H \varphi^2\right]\left\{\frac{2}{\epsilon}
+\ln{\frac{4\pi\mu^2}{M^2(0)}}-\gamma\right\}\; .
\end{eqnarray}
The term $\lambda(\xi-1/6)R\varphi^2$ in (\ref{fdiv})
 is cancelled by the
same term with opposite sign in (\ref{sdiv}).
The $H^2\varphi^2$-terms in (\ref{sdiv}) and (\ref{tdiv})
are absorbed by the counter term proportional to
$\delta \xi$, and the divergence
proportional to
$\varphi\varphi'$ is compensated by this counter term as well.
The remaining $\varphi$-independent but still time-dependent divergent
terms
are absorbed into the counter terms $\delta\tilde{\alpha} ~H_{tt}$ 
and $\delta\tilde{ Z}~ G_{tt}$.
We choose
\begin{eqnarray}
\delta \tilde{\alpha} &=&
-\frac{(\xi-{\textstyle \frac{1}{6}})^2}{32\pi^2}
\left\{\frac{2}{\epsilon}+
\ln{\frac{4\pi\mu^2}{m^2}}-\gamma\right\}\; , \\
\delta\tilde {Z}& =& - \frac{(\xi-{\textstyle \frac{1}{6}} ) m^2}{16\pi^2}
\left\{\frac{2}{\epsilon}+
\ln{\frac{4\pi\mu^2}{m^2}}-\gamma\right\} 
\; .
\end{eqnarray}
That is, all divergent integrals appearing in the unrenormalized
divergent
fluctuation integral of the energy are removed by the corresponding
counter terms
and we finally arrive at the renormalized expression for the energy
\begin{eqnarray}
{\cal E}_{ren}&=&
\frac{1}{2a^4}\varphi'^2+\frac{1}{2a^2} (m^2+\Delta m^2)\varphi^2
+\frac{\lambda+\Delta \lambda}{4!a^4}\varphi^4
-6\left(\xi-{\textstyle \frac{1}{6}}+\Delta \xi\right)\left(
\frac{H^2}{2a^2}\varphi^2-\frac{H}{a^3}
\varphi\varphi'
\right)
\nonumber\\
&&+\Delta \tilde{\Lambda}+\Delta \tilde{\alpha} H_{tt}
+\Delta\tilde{Z} G_{tt}+\frac{M^2(0)}{32\pi^2a^2}(6\xi-1)
\left(\frac {R}{ 6}+H^2\right)
+\frac{M^4(0)}{128\pi^2a^4}-\frac{M^2(0)M^2(\tau)}{32\pi^2a^4}
\nonumber\\
&&+\frac{1}{a^2}\int\! \frac{d^3k}{(2\pi)^3}\Biggl\{
\frac{V(\tau)}{4\Omega_{k0}a^2}
\left(2{\rm Re }\,{h}_k^{(\overline{2})}+|h_k^{(\overline{1})}|^2\right)
+\frac{1}{4\Omega_{k0}a^2}|{h_k'}^{(\overline{2})}|^2
\nonumber\\
&&\hspace{2.5cm}
+\frac{1}{2\Omega_{k0}a^2}{\rm Re }\,\left({h_k'}^{(1)}
{h_k'}^{(\overline{2})*}\right)
+\frac{1}{16\Omega_{k0}^3a^2}{\cal C}^2(V',\tau)
+\frac{1}{16\Omega_{k0}^3a^2}{\cal S}^2(V',\tau)
\nonumber\\
&&\hspace{2.5cm}-\frac{1}{4\Omega_{k0}}(6\xi-1)
\left(\frac {R}{6}+H^2\right)\left[
\frac{1}{2\Omega_{k0}^2}{\cal C}(V',\tau)
+2{\rm Re }\,{h_k}^{(\overline{2})}+|h_k^{(\overline{1})}|^2\right]
\nonumber\\
&&\hspace{2.5cm}+\frac{1}{4\Omega_{k0}} (6\xi-1)\frac {H}{a}\biggl[
\frac{1}{2\Omega_{k0}^2}{\cal C}(V'',\tau')
+\frac{1}{2\Omega_{k0}^2}\cos{2\Omega_{k0}\tau}V'(0)\nonumber\\ 
&&\hspace{5.8cm}
+2{\rm Re }\, h_k^{(\overline{2})'}
+2{\rm Re }\, h_k^{(\overline{1})*}h_k^{(\overline{1})'}\biggr]	\Biggr\}
\; .
\end{eqnarray}
The finite corrections are
\begin{eqnarray}
\Delta \tilde{\alpha} &=&
\frac{\left(\xi-{\textstyle \frac{1}{6}}\right)^2}{32\pi^2}\ln{\frac{m^2}{M^2(0)}}\; ,
\\
\Delta \tilde{\Lambda} &=&-\frac{m^4}{64\pi^2}\ln{\frac{m^2}{M^2(0)}}\; ,
\\
\Delta \tilde{Z}& =&
\frac{\left(\xi-{\textstyle \frac{1}{6}}\right)^2m^2}{16\pi^2}\ln{\frac{m^2}{M^2(0)}}\; .
\end{eqnarray} 

Next we have to consider the renormalization of the
trace of the energy momentum tensor.
We introduce the available counter terms into the unrenormalized
expression for
$T_{\mu}^\mu$ so that it reads now
\begin{eqnarray}
T_{\mu}^\mu&=&
-\left[1-6(\xi+\delta \xi)\right]\left(
\frac{\varphi'^2}{a^4}+ \frac{H^2}{a^2}\varphi^2
-2\frac{H}{a^3}\varphi\varphi'
\right)+\frac{6(\xi+\delta\xi)}{a^4}\varphi\varphi''\nonumber\\
&&+2(m^2+\delta m^2)
\frac{\varphi^2}{a^2}+\frac{\lambda+\delta \lambda}{6a^4}
\varphi^4
+4\delta \tilde{\Lambda}+
\delta \tilde{\alpha}  H^{\mu}_\mu + \delta\tilde{Z} G_\mu^\mu
\nonumber\\
&&+\frac{1}{a^2}\int\! \frac{d^3k}{(2\pi)^3}\Biggl\{\frac{1}{a^2}(1-6\xi)
\Biggl[-\frac{1}{2\Omega_{k0}}|{h'_k}^{(\overline{1})}|^2
+\frac{i}{2}\left(-2i{\rm Im }\,{h'_k}^{
(\overline{1})}-2i
{\rm Im }\, h_k^{(\overline{1})*}{h'_k}^{(\overline{1})}\right)
\nonumber\\
&&\hspace{4.5cm}+\frac{1}{2\Omega_{k0}}V(\tau)
\left(1+2{\rm Re }\,{h_k}^{(\overline{1})}+|h_k^{(\overline{1})}|^2\right)\Biggr]
\nonumber\\
&&\hspace{2.5cm}+\frac{H}{2\Omega_{k0}a}(1-6\xi)\frac{d}{d\tau}\left(1
+2{\rm Re }\,{h_k}^{(\overline{1})}+|h_k^{(\overline{1})}|^2\right)
\nonumber\\
&&\hspace{2.5cm}+
\frac{1}{2\Omega_{k0}}\left[m^2+\frac{\lambda}{2a^2}\varphi^2-(1-6\xi)
\left(H^2-\frac {R}{6}\right)\right]
\nonumber\\
&&\hspace{3cm}\times\left.
\left(1+2{\rm Re }\,{h_k}^{(\overline{1})}+|h_k^{(\overline{1})}|^2\right)
\right.\Biggr\}\; .\end{eqnarray}
We can split the trace of
the stress tensor into a divergent and into a convergent
part.
The divergent part reads
\begin{eqnarray}\label{tmumudiv}
{T_{\mu}^{\mu}}_{div}&=&\frac{1}{a^2}\int\! \frac{d^3k}{(2\pi)^3}\Biggl\{
\frac{1}{a^2}(1-6\xi)\frac{1}{8(\Omega_{k0})^3}V''(\tau)
-(1-6\xi)\frac{H}{a}\frac{1}{4(\Omega_{k0})^3}V'(\tau)
\nonumber\\
&&
\hspace{2.5cm}+\left[m^2+\frac{\lambda}{2a^2}\varphi^2-(1-6\xi)
(H^2-{\textstyle \frac{1}{6}} R)\right]
\left[\frac{1}{2\Omega_{k0}}-\frac{1}{4(\Omega_{k0})^3}V(\tau)\right]
\Biggr\}\; .\end{eqnarray}
The first derivative of the potential cancels 
with the corresponding term from (\ref{vsecder})
in (\ref{tmumudiv}). The remaining divergent parts are
\begin{eqnarray}
{T_{\mu}^{\mu}}_{div}&=&\frac{1}{a^2}\int\! \frac{d^3k}{(2\pi)^3}
\Biggl\{(1-6\xi)\frac{1}{8(\Omega_{k0})^3}
\Biggl[
{\textstyle {\textstyle \frac{1}{3}}}  R M^2(\tau)-2  H^2 M^2(\tau)\nonumber\\
&&\hspace{5.2cm}+2 aH\left(\xi-{\textstyle \frac{1}{6}} \right)R'-2\lambda \frac H a
\varphi\varphi'+\left(\xi-{\textstyle \frac{1}{6}}\right)R''
\nonumber\\
&&\hspace{5.2cm}+\frac{1}{a^2}\lambda(\varphi\varphi''+\varphi'^2)
-\lambda{\textstyle \frac{1}{6}}R\varphi^2 +\lambda  H^2\varphi^2\Biggr]
\nonumber\\
&&\hspace{1.5cm}
+\left[m^2+\frac
{\lambda}{2a^2}\varphi^2-(1-6\xi)\left(H^2-\frac R 6\right)\right]
\left[\frac{1}{2\Omega_{k0}}-\frac{1}{4(\Omega_{k0})^3}V(\tau)\right]
\Biggr\}
\; .\end{eqnarray}
After dimensional regularization again the counter terms 
absorb all divergent
terms in the fluctuation integral of $T_{\mu}^\mu$. 
The renormalized trace of the stress tensor takes
the final form
\begin{eqnarray}
T_{\mu}^\mu&=&
-\left[1-6(\xi+\Delta \xi)\right]\left(
\frac{\varphi'^2}{a^4}+ 
\frac{H^2}{a^2}\varphi^2-2\frac{H}{a^3}\varphi\varphi'
\right)+\frac{6(\xi+\Delta \xi)}{a^4}\varphi\varphi''\nonumber\\
&&+2(m^2+\Delta m^2)
\frac{\varphi^2}{a^2}+\frac{\lambda+\Delta \lambda}{6a^4}
\varphi^4
+4\Delta \tilde{\Lambda}+
\Delta \tilde{\alpha} H_{\mu}^{\mu} + \Delta \tilde{Z} G_\mu^\mu 
\nonumber \\
&&-\frac{M^2(0)}{16\pi^2a^2}\left[m^2+\frac{\lambda}{2a^2}\varphi^2
-(1-6\xi)\left(H^2-\frac R 6\right)\right]
\nonumber\\
&&+\frac{1}{a^2}\int\! \frac{d^3k}{(2\pi)^3}\Biggl\{\frac{1}{a^2}(1-6\xi)
\Biggl[-\frac{1}{2\Omega_{k0}}|{h'_k}^{(\overline{1})}|^2
+\frac{i}{2}\left(-2i{\rm Im }\, h_k^{\prime(\overline{2})}-2i
{\rm Im }\, h_k^{(\overline{1})*}h_k^{\prime(\overline{1})}\right)
\nonumber\\
&&\hspace{4.7cm}+\frac{1}{2\Omega_{k0}}V(\tau)
\left(2{\rm Re }\,{h_k}^{(\overline{1})}+|h_k^{(\overline{1})}|^2\right)
+\frac{1}{4\Omega_{k0}^2}V'(0)\sin 2\Omega_{k0}\tau\nonumber\\
&&\hspace{4.7cm}-\frac{1}{8\Omega_{k0}^3}V''(0)\cos 2\Omega_{k0}\tau
-\frac{1}{8\Omega^3_{k0}}{\cal C}(V''',\tau)
\Biggr]\nonumber\\
&&\hspace{2.5cm}+\frac{H}{2\Omega_{k0}a}(1-6\xi)\Biggl[
\frac{1}{2\Omega_{k0}^2}{\cal C}(V'',\tau)
+2{\rm Re }\, h_k^{\prime(\overline{2})}\nonumber\\
&&\hspace{5.5cm}+\frac{1}{2\Omega_{k0}^2}\cos{2\Omega_{k0}\tau}V'(0)
+2{\rm Re }\,  h_k^{\prime(\overline{1})*}h_k^{(\overline{1})}\Biggr]\nonumber\\
&&\hspace{2.5cm}+
\frac{1}{2\Omega_{k0}}\left[m^2+\frac{\lambda}{2a^2}\varphi^2
-(\xi-{\textstyle \frac{1}{6}})\left(H^2-\frac R 6\right)
\right]
\nonumber\\
&&\hspace{4.5cm}\times
\left[
\frac{1}{2\Omega_{k0}^2}{\cal C}(V',\tau)
+2{\rm Re }\,{h_k}^{(\overline{2})}+|h_k^{(\overline{1})}|^2\right]
\Biggr\}
\; .\end{eqnarray}
\section{Conclusions}
We have presented here the renormalized equations of motion for
a scalar field in a conformally flat FRW universe including
the quantum back reaction in one-loop approximation.
We have used dimensional regularization and an $\overline{MS}$
renormalization. However, within this formalism 
other renormalization conditions can be employed, as well
as other regularizations.
The formalism is fully covariant and the counter terms
can be chosen independent of the initial conditions.
In contrast to the adiabatic regularization it is not based on
the WKB expansion and can therefore be generalized to coupled
systems \cite{Baacke:1997b}. Furthermore, we do not have to
perform delicate subtractions in the divergent integrals since
in our formulation they are finite from the outset.
The method can be adapted easily to finite
temperature computations as well, since no new ultraviolet divergencies
are introduced. 

The intention of this work is of course to supply 
at the same time a useful scheme for numerical computations. 
We have shown previously \cite{Baacke:1997a,Baacke:1997b}
that the numerical implementation is straightforward.
A realistic application to one of the inflationary scenarios
needs, however, a judicious choice of initial conditions
and of the parameters of the theory, as well as
extensive numerical experiments. 
While such computations are in progress we prefer to
present here the formalism as such; we
think that is by itself of general theoretical 
as well as of practical interest.
\section*{Acknowledgements}
It is a pleasure to thank Andreas Ringwald and
Daniel Boyanovsky for useful and inspiring discussions.

\end{document}